\documentclass[11pt]{article}

\usepackage[T1]{fontenc}
\usepackage[utf8]{inputenc}
\usepackage{lmodern}
\usepackage[a4paper,margin=1in]{geometry}
\usepackage{authblk}
\usepackage{amsmath}
\usepackage{amssymb}
\usepackage{bm}
\usepackage{booktabs}
\usepackage{graphicx}
\usepackage{pgf}
\usepackage{tikz}
\usepackage{microtype}
\usepackage[numbers,sort&compress]{natbib}
\usepackage[hidelinks]{hyperref}

\graphicspath{{./pics/}}
\begin{document}

\title{Accelerated ``on-the-fly'' coupled-cluster path-integral molecular
dynamics: Impact of nuclear quantum effects on an asymmetric proton}

\author[1]{Thomas Spura}
\author[1]{Hossam Elgabarty}
\author[2,3,4]{Thomas D. K\"uhne}
\affil[1]{Dynamics of Condensed Matter and Center for Sustainable Systems Design, Chair of Theoretical Chemistry, University of Paderborn, Warburger Str. 100, D-33098 Paderborn, Germany}
\affil[2]{Center for Advanced Systems Understanding (CASUS), Conrad-Schiedt-Stra{\ss}e 20, 02826 G\"orlitz, Germany}
\affil[3]{Helmholtz Zentrum Dresden-Rossendorf, Bautzner Landstra{\ss}e 400, 01328 Dresden, Germany}
\affil[4]{Institute of Artificial Intelligence, Technische Universit\"at Dresden, Helmholtzstra{\ss}e 10, 01069 Dresden, Germany}
\date{}

\maketitle

\begin{abstract}
We present an accelerated ``on-the-fly'' coupled-cluster path-integral molecular dynamics (PIMD) method for finite-temperature simulations in which electron correlation and nuclear quantum effects are treated simultaneously. The approach is based on our quantum ring-polymer contraction (qRPC) technique, in which the inexpensive Hartree-Fock potential is evaluated on the full ring-polymer, while the expensive coupled-cluster correction is evaluated on the centroid only. This qRPC decomposition is combined with a second-generation Car-Parrinello-like dynamics of the Hartree-Fock reference and a basis-consistent extrapolation of the coupled-cluster and de-excitation amplitudes. The combination of all three acceleration layers is essential for making correlated PIMD calculations feasible. We apply the method to a proton shared by water and formaldehyde. Relative to classical nuclei, nuclear quantum effects broaden covalent X--H bond-length distributions, reduce the bias of the shared proton toward formaldehyde, and shift the mean proton-transfer coordinate from 0.206 to 0.135~\AA\@. The probability of finding the proton closer to formaldehyde decreases from 81.7\% to 61.1\%. The corresponding nuclear magnetic shielding tensors show that electron correlation and nuclear quantum effects are of comparable magnitude and can act in opposite directions. These results demonstrate that predictive simulations of asymmetric hydrogen bonds require a simultaneous treatment of correlated electronic structure and nuclear quantum fluctuations.
\end{abstract}

\noindent\textbf{Correspondence:} Thomas D. K\"uhne, \href{mailto:tkuehne@cp2k.org}{tkuehne@cp2k.org}


\medskip
\noindent\textbf{Abbreviations:} CC, coupled cluster; CC-PIMD, coupled-cluster path-integral molecular dynamics; CCSD, coupled cluster with single and double excitations; DFT, density-functional theory; HF, Hartree--Fock; MD, molecular dynamics; NMR, nuclear magnetic resonance; NQE, nuclear quantum effect; PCF, pair correlation function; PIMD, path-integral molecular dynamics; qRPC, quantum ring-polymer contraction.

\section{Introduction}

Nuclear quantum effects (NQEs), including zero-point motion and tunneling, are essential for a quantitative description of systems containing light atoms or low temperatures. As such, they are particularly important for hydrogen-bonded systems, proton-transfer reactions, and aqueous environments, where the quantum delocalization of the proton can change both structure and spectroscopy. Path-integral molecular dynamics (PIMD) provides a practical route to quantum thermal averages by mapping each quantum particle onto a classical ring-polymer~\cite{Feynman,parrinello1984moltenkcl,chandler1981isomorphism, Habershon2013}. 

The main practical bottleneck in \textit{ab initio} PIMD is the repeated evaluation of forces for all beads of the ring-polymer. Density-functional theory (DFT) is therefore the electronic-structure method most commonly used in ``on-the-fly'' PIMD simulations~\cite{marx1994pimd,tuckerman1996picpmd,Voth1998,shiga2001aimd}. DFT-based PIMD has successfully described quantum fluctuations in water, ice, proton-transfer systems, and isotope-dependent hydrogen bonds~\cite{Tuckerman1997,Kohanoff1998,Benoit1998,Marx1999,Tuckerman2002,PhysRevLett.101.017801,kaczmarek2009onthefly,PhysRevLett.104.066102,Li2011,Ceriotti2013,Giberti2014,kessler2015structure,Spura_2014,Singh2018,Clark2019NQEHBond}. However, even small electronic-structure errors are critical on the Kelvin energy scale. An error of only 0.2~kcal~mol$^{-1}$, about an order of magnitude smaller than the strength of a typical hydrogen bond in water~\cite{kuehnewater2013,kuehnewater2014}, can qualitatively affect a finite-temperature simulation~\cite{kuehnewater2009}.

Wave-function-based electronic-structure methods offer a more systematic treatment of electron correlation~\cite{PopleNobel,HelgakerBook}, but their high cost has so far limited their use in molecular dynamics (MD) and especially in PIMD\@. In an earlier work, ``on-the-fly'' coupled-cluster PIMD (CC-PIMD) was introduced for the protonated water dimer~\cite{spura2015ccpimd}. Here, we extend this strategy to an asymmetric proton shared by water and formaldehyde and focus on the acceleration required to make such simulations practical. The central ingredient is quantum ring-polymer contraction (qRPC)~\cite{john2015qrc}, which extends the ring-polymer contraction idea of Manolopoulos and coworkers~\cite{markland2008refinedcontraction,comp_quant_eff}. In the present coupled-cluster (CC) setting, the natural auxiliary potential is Hartree-Fock (HF), because it is already required to construct the CC reference determinant. The costly residual potential is then the difference between CC singles and doubles (CCSD) and HF\@.

The qRPC decomposition reduces the number of expensive correlated evaluations along the imaginary-time ring-polymer, but it does not by itself remove the cost of the individual HF and CCSD calculations. We therefore combine qRPC with two fictitious second-generation Car-Parrinello-like dynamics~\cite{kuehne2007,CP2Greview}. The HF auxiliary potential is accelerated by a density-matrix predictor, while the CCSD residual is accelerated by amplitude transformation and extrapolation. The practical method is the combination of these imaginary-time and real-time acceleration layers; without that combination, on-the-fly CC-PIMD with statistically meaningful trajectories would remain prohibitively expensive.

The remainder of this manuscript is organized as follows. Section~II summarizes the CC formalism needed for force calculations. Section~III describes the accelerated CC-PIMD scheme, starting from the qRPC acceleration strategy and then detailing the HF and CCSD predictors. Section~IV presents the acceleration benchmarks and the impact of NQEs on the structure and nuclear magnetic shielding tensors of the asymmetric proton. Section~V summarizes the main conclusions.

\section{Coupled-Cluster Theory}

Within CC theory, the correlated wave function is represented by the exponential ansatz~\cite{civzek1966correlationproblem,Gauss1998,HelgakerBook,bartlett2007CCreview}
\begin{equation}
    |\Psi_{\text{exact}}\rangle = e^{\hat{T}} |\Psi\rangle,
\label{eq:CC}
\end{equation}
where the reference wave function $|\Psi\rangle$ is a Slater determinant built from HF molecular orbitals $|\psi_i\rangle$. The exponential form ensures size extensivity, so that the correlation energy scales correctly with system size for separated fragments~\cite{Hanrath_2009,Helgaker1999,Bartlett_1978}. The cluster operator is expanded as
\begin{equation}
    \hat{T} = \hat{T}_1 + \hat{T}_2 + \hat{T}_3 + \ldots,
\label{eq:T_expansion}
\end{equation}
where the $n$-tuple excitation operator is
\begin{equation}
  \hat{T}_n = \frac{1}{{(n!)}^2}
  \sum_{i,j,k,\ldots}\sum_{a,b,c,\ldots}
  t_{ijk\cdots}^{abc\cdots}
  \hat{c}_a^{\dag}\hat{c}_b^{\dag}\hat{c}_c^{\dag}\cdots
  \hat{c}_i\hat{c}_j\hat{c}_k.
\end{equation}
Indices $i,j,k,\ldots$ denote occupied orbitals and $a,b,c,\ldots$ denote virtual orbitals. The operators $\hat{c}^{\dag}$ and $\hat{c}$ are creation and annihilation operators, respectively. The cluster amplitudes $t_{ijk\cdots}^{abc\cdots}$ are obtained from the coupled nonlinear amplitude equations. In the limit of all excitations and a complete one-particle basis, the ansatz converges to the exact wave function.

Because the usual CC energy expression is not stationary with respect to the cluster amplitudes, analytic forces and other energy derivatives are formulated using the Lagrangian functional~\cite{Adamowicz_1984,Koch_1990,Szalay_1995,Arponen_1983}
\begin{equation}
    \tilde{E} = \langle\Psi|(1+\hat{\Lambda})e^{-\hat{T}}\hat{H}e^{\hat{T}}|\Psi\rangle,
\label{eq:functional}
\end{equation}
where the de-excitation operator has the expansion
\begin{equation}
    \hat{\Lambda} = \hat{\Lambda}_1 + \hat{\Lambda}_2 + \hat{\Lambda}_3 + \ldots.
\label{eq:lambda_expansion}
\end{equation}
In second quantization,
\begin{equation}
    \hat{\Lambda}_n = \frac{1}{{(n!)}^2}
    \sum_{i,j,k,\ldots}\sum_{a,b,c,\ldots}
    \lambda_{abc\cdots}^{ijk\cdots}
    \hat{c}_i^{\dag}\hat{c}_j^{\dag}\hat{c}_k^{\dag}\cdots
    \hat{c}_a\hat{c}_b\hat{c}_c.
\end{equation}
The functional in Eq.~\ref{eq:functional} is stationary with respect to the $\lambda$ amplitudes at the CC solution and is therefore suitable for evaluating energy derivatives~\cite{Kallay2003AnalyticDerivatives}. Throughout this work, the CCSD approximation is used for MD/PIMD, i.e., $\hat{T}$ and $\hat{\Lambda}$ are truncated after single and double excitations.

\section{Accelerated Coupled-Cluster Path-Integral Molecular Dynamics}

\subsection{qRPC-guided acceleration strategy}

The primary bottleneck in correlated PIMD is the bead multiplication of the electronic-structure force evaluation. The organizing idea of the present method is therefore qRPC: the full ring-polymer is described by an inexpensive auxiliary potential, whereas only a smoother and more expensive correction is evaluated on a contracted ring-polymer. For CCSD, the natural auxiliary potential is HF, because the HF reference determinant must be generated for every CCSD calculation anyway. The expensive correction is the CCSD--HF residual.

This qRPC separation defines the imaginary-time acceleration, but the two electronic-structure pieces still need to be evaluated efficiently along the real-time MD/PIMD trajectory. The HF reference and the CCSD amplitude equations both vary smoothly from one time step to the next and can therefore be accelerated by second-generation Car-Parrinello-like predictors that use information from previous steps. In the complete algorithm, qRPC reduces the number of expensive bead evaluations, the HF predictor accelerates the auxiliary potential on the full ring-polymer, and the amplitude predictor accelerates the contracted CCSD residual. These components are complementary rather than interchangeable.

\subsection{HF auxiliary potential}

For the HF part, we avoid recomputing the reference determinant from scratch at every \textit{ab initio} MD/PIMD time step. Following the second-generation Car-Parrinello (CP2G) strategy~\cite{kuehne2007,CP2Greview} and our previous CC-MD/PIMD work~\cite{spura2015ccpimd}, we predict the occupied subspace from the single-particle density operator $\hat{\rho}=\sum_i|\psi_i\rangle\langle\psi_i|$, which is smoother than the individual molecular orbitals. At time $t_n$, the predicted orbitals are obtained from the occupied subspace at the previous time step via
\begin{eqnarray}
|\psi_i^p(t_n)\rangle \approx
\underbrace{\sum_{m=1}^K (-1)^{m+1}m
\frac{\binom{2K}{K-m}}{\binom{2K-2}{K-1}}
\hat{\rho}(t_{n-m})}_{\hat{\rho}^p(t_n)}
|\psi_i(t_{n-1})\rangle.
\label{eq:hf_predictor}
\end{eqnarray}
Here, $K$ is the predictor length. The resulting orbitals are used as an initial guess for the HF problem and remain close to the instantaneous Born-Oppenheimer surface.

\subsection{CCSD residual potential}

The dominant cost of a CCSD force calculation is the iterative solution of the cluster and $\hat{\Lambda}$ amplitude equations. Reusing amplitudes from previous geometries is not sufficient by itself, because the molecular orbital basis can change by arbitrary unitary transformations without changing the HF energy. These gauge-like changes introduce spurious variations in the amplitudes. To remove this nonphysical contribution, the amplitudes are first transformed to an intermediate representation in the full symmetrically orthogonalized atomic-orbital basis,
\begin{equation}
    \tilde{C}_{\mu p} =
    \sum_{\gamma}^{n_{\text{ao}}}
    (S^{1/2})_{\mu\gamma} C_{\gamma p},
    \qquad
    S_{\mu\gamma}=\langle\phi_{\mu}|\phi_{\gamma}\rangle,
\label{eq:orthogonalized_coefficients}
\end{equation}
where $p$ denotes a molecular orbital and $\phi_{\mu}$ denotes an atomic basis function~\cite{Lo_wdin_1950,Slater_1954}. The Löwdin basis functions themselves are generated with $S^{-1/2}$; Eq.~\ref{eq:orthogonalized_coefficients} uses $S^{1/2}$ because it gives the corresponding molecular-orbital coefficients in that orthonormal basis. For clarity, the following equations are written for the one-body cluster amplitudes; higher-body amplitudes are transformed analogously for each occupied and virtual index. At time $t-\Delta t$, the transformation to the orthogonal basis is
\begin{equation}
    \tilde{t}_{\mu}^{\nu,t-\Delta t} =
    \sum_{ia}
    \tilde{C}_{\mu i}^{t-\Delta t}
    t_i^{a,t-\Delta t}
    \tilde{C}_{\nu a}^{t-\Delta t}.
\label{eq:transformation}
\end{equation}
The back-transformation to the molecular orbital basis at the new geometry is
\begin{equation}
    t_i^{a,t} =
    \sum_{\mu\nu}
    \tilde{C}_{\mu i}^{t}
    \tilde{t}_{\mu}^{\nu,t}
    \tilde{C}_{\nu a}^{t}.
\label{eq:transformation_back}
\end{equation}
In the intermediate representation, the physical evolution of the amplitudes can be extrapolated to the next time step using the always-stable predictor of Kolafa~\cite{kolafa2004aspc},
\begin{equation}
    \tilde{t}^{\nu,t}_{\mu} \approx
    \sum_{m=1}^K (-1)^{m+1}m
    \frac{\binom{2K}{K-m}}{\binom{2K-2}{K-1}}
    \tilde{t}^{\nu,t-m\Delta t}_{\mu}.
\label{eq:t_extrapol}
\end{equation}
The same transformation and extrapolation are applied to the $\hat{\Lambda}$ amplitudes. Since only standard matrix multiplications are involved, the formal scaling of CCSD is not changed.

\subsection{Combined qRPC/CP2G force evaluation}

With the HF and CCSD real-time predictors in place, the path-integral bead contraction is applied to the CCSD/HF decomposition. For a ring-polymer with $P$ beads, the CCSD potential averaged over the full ring-polymer can be decomposed exactly as
\begin{equation}
V_P^{\mathrm{CCSD}}(\mathbf{R}) =
\frac{1}{P}\sum_{s=1}^{P} V_{\mathrm{HF}}(\mathbf{R}^{(s)})
+ \frac{1}{P}\sum_{s=1}^{P} \Delta V_{\mathrm{CCSD}}(\mathbf{R}^{(s)}),
\label{eq:qrpc_decomposition}
\end{equation}
with
\begin{equation}
\Delta V_{\mathrm{CCSD}}(\mathbf{R}) =
V_{\mathrm{CCSD}}(\mathbf{R}) - V_{\mathrm{HF}}(\mathbf{R}).
\label{eq:ccsd_residual}
\end{equation}
The HF term is the inexpensive auxiliary potential and is evaluated on every bead. This is advantageous in the present method because the HF reference is required anyway for each CCSD force calculation and is accelerated by Eq.~\ref{eq:hf_predictor}. The residual $\Delta V_{\mathrm{CCSD}}$ is the expensive but smoother part of the potential and can be evaluated on a contracted ring-polymer with $P'<P$ beads,
\begin{equation}
V_P^{\mathrm{CCSD}}(\mathbf{R}) \approx
\frac{1}{P}\sum_{s=1}^{P} V_{\mathrm{HF}}(\mathbf{R}^{(s)})
+ \frac{1}{P'}\sum_{s'=1}^{P'}
\Delta V_{\mathrm{CCSD}}(\tilde{\mathbf{R}}^{(s')}).
\label{eq:qrpc_ccsd}
\end{equation}
The contracted coordinates $\tilde{\mathbf{R}}^{(s')}$ are obtained by the standard normal-mode contraction of the full ring-polymer~\cite{markland2008efficientcontraction,markland2008refinedcontraction}. The corresponding forces follow by differentiating Eq.~\ref{eq:qrpc_ccsd} and back-transforming the contracted-ring-polymer forces to the full bead representation. Thus, the inexpensive HF bead forces sample the high-frequency ring-polymer fluctuations and are accelerated by Eq.~\ref{eq:hf_predictor}, while the expensive CCSD correction is accelerated both in imaginary time by qRPC and in real time by Eqs.~\ref{eq:transformation}--\ref{eq:t_extrapol}. This combined qRPC/CP2G construction is what makes the on-the-fly correlated MD/PIMD simulation computationally viable. In practice, the contracted ring-polymer can \textit{de facto} be reduced to the centroid, i.e., $P'=1$, because the CCSD--HF residual is smooth with respect to the internal ring-polymer modes.

    \begin{figure}[htbp]
        \centering
        \resizebox{0.95\textwidth}{!}{\input{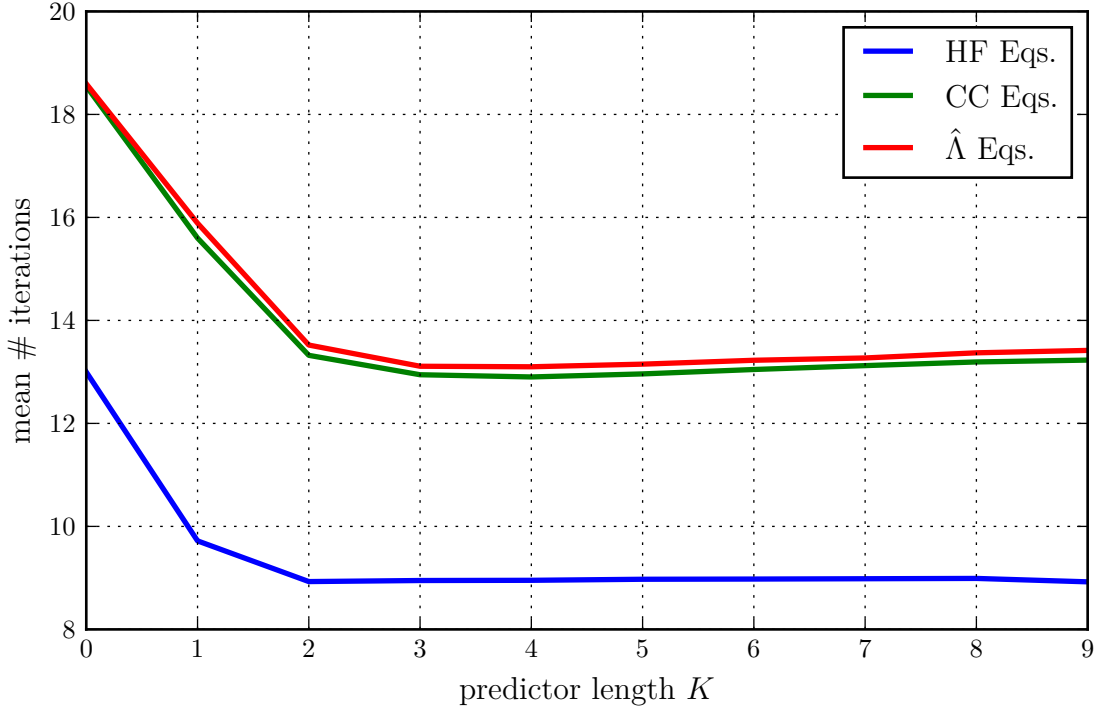}}
        \caption{Mean number of iterations required to converge the cluster and $\hat{\Lambda}$ equations as a function of the predictor length $K$ used in the amplitude extrapolation.}
        \label{fig:ASYM_speedup}
    \end{figure}

\section{Results}

\subsection{Computational details}

The CC-MD and CC-PIMD simulations were performed with i-PI~\cite{ipi2014}. Contrary to our previous work~\cite{spura2015ccpimd}, where CFOUR~\cite{Matthews2020CFOUR} had been used, forces were evaluated at the CCSD/cc-pVDZ level of theory~\cite{Scheiner1987,Stanton1991,gauss1991gradients,Gauss1991,Dunning1989} using a modified version of the PSI4 software package~\cite{PSI4_2011,Crawford2000} that includes the amplitude transformation and extrapolation described above. All simulations were carried out in the canonical constant-number, constant-volume, and constant-temperature (NVT) ensemble at 300~K with a time step of 0.25~fs. The production trajectories were 25~ps long, which in total corresponds to 200000 CCSD force calculations. The PIMD simulations used $P=32$ beads, and a corresponding classical-nuclei CC-MD simulation was performed with $P=1$.

The cc-pVDZ basis used for the on-the-fly CCSD dynamics should not be regarded as sufficient for basis-set-converged CC energies or absolute spectroscopic observables. A larger basis would be desirable for quantitative benchmark calculations, but is not computationally feasible for the correlated MD/PIMD trajectories considered here. Because the CC-MD and CC-PIMD trajectories are generated consistently at the same electronic-structure level, the chosen basis is sufficient for the present purpose of qualitatively quantifying the influence of NQEs.

Nuclear magnetic shielding tensors were evaluated as ensemble averages over 1000 decorrelated snapshots from the CC-MD and CC-PIMD trajectories, separated by 25~fs each. These calculations were performed at the CCSD(T)/cc-pVTZ level of theory, where CCSD(T) denotes CCSD with perturbative triples~\cite{GIAO_CCSD_T_Gauss1996,GaussStantonNMR2004}, using CFOUR~\cite{Harding2008,Matthews2020CFOUR}. HF shielding tensors were computed for the same snapshots to separate the effects of electron correlation from those of nuclear quantum fluctuations.

\subsection{Acceleration of the coupled-cluster equations}

The amplitude predictor reduces the number of CCSD and $\hat{\Lambda}$ iterations substantially. Even when the direct inversion in the iterative subspace (DIIS) convergence accelerator is used~\cite{Pulay_1980,Pulay_1982}, Fig.~\ref{fig:ASYM_speedup} shows that the number of iterations required to reach the electronic ground state decreases by about 30\%. The HF part also converges in fewer iterations because of the density-matrix predictor in Eq.~\ref{eq:hf_predictor}. In a CCSD gradient calculation, however, the HF part accounts for only a small fraction of the total cost, so the practical speed-up is governed primarily by the reduction in cluster and $\hat{\Lambda}$ iterations.

\begin{figure}[htbp]
        \centering
        \resizebox{0.95\textwidth}{!}{\input{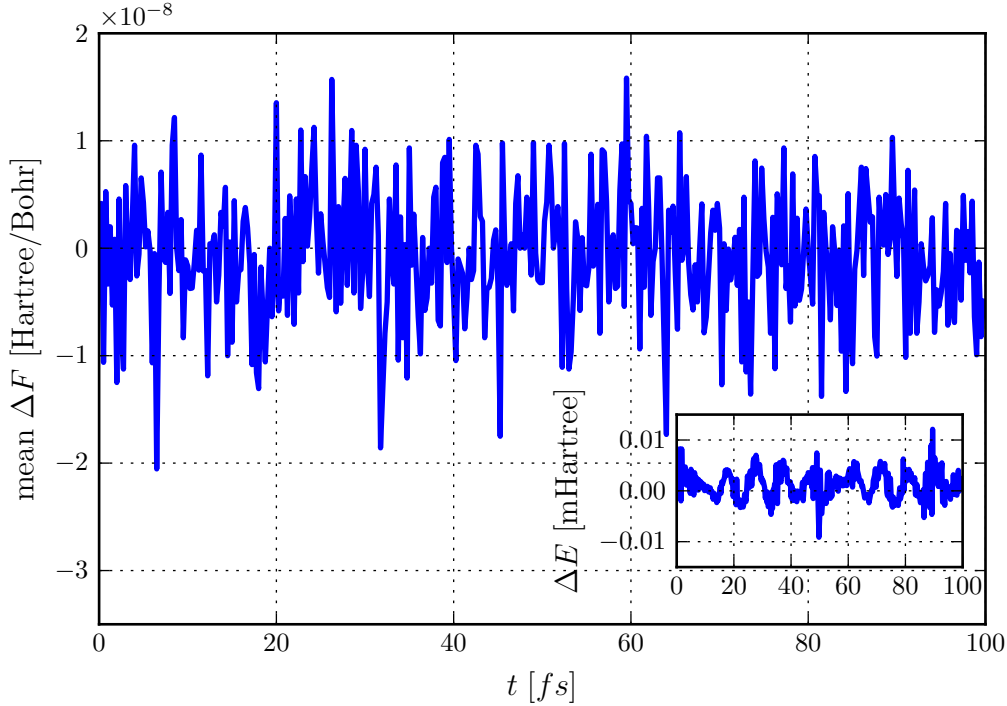}}
        \caption{Energy and force deviations along a CC-MD trajectory propagated with two iterations of the cluster equations and fully converged $\hat{\Lambda}$ equations, relative to a retraced trajectory with fully converged CCSD and $\hat{\Lambda}$ amplitudes.}
        \label{fig:ASYM_reftraj}
    \end{figure}

We also tested whether the CCSD amplitudes need to be fully converged at every time step. A short CC-MD simulation was propagated using only two iterations for the cluster amplitudes, while the $\hat{\Lambda}$ equations were fully converged. The same nuclear trajectory was then retraced with fully converged cluster and $\hat{\Lambda}$ amplitudes. The resulting energy and force differences are shown in Fig.~\ref{fig:ASYM_reftraj}. The energy error remains below $10^{-5}$~Hartree, with an average of $1.13\times10^{-6}$~Hartree. The mean signed force difference is $-7.16\times10^{-10}$~Hartree~\AA$^{-1}$ and the maximum absolute component is $1.58\times10^{-8}$~Hartree~\AA$^{-1}$. These small deviations indicate that a small, fixed number of cluster-amplitude iterations can provide a controlled approximation for CC-MD/PIMD once a high-quality predictor is available.

Since the force deviation can be considered to a sufficiently high degree as white, the computational effort could be reduced further by replacing strictly Hamiltonian dynamics with a modified Langevin equation. In principle, the latter allows to rigorously compensate the effective noise introduced by the incomplete convergence of the cluster equations, thereby ensuring an exact sampling of the Boltzmann distribution in the spirit of CP2G~\cite{kuehne2007,CP2Greview}.

\subsection{Nuclear quantum effects on the asymmetric proton}

\begin{figure}[htbp]
        \centering
        \resizebox{0.95\textwidth}{!}{\input{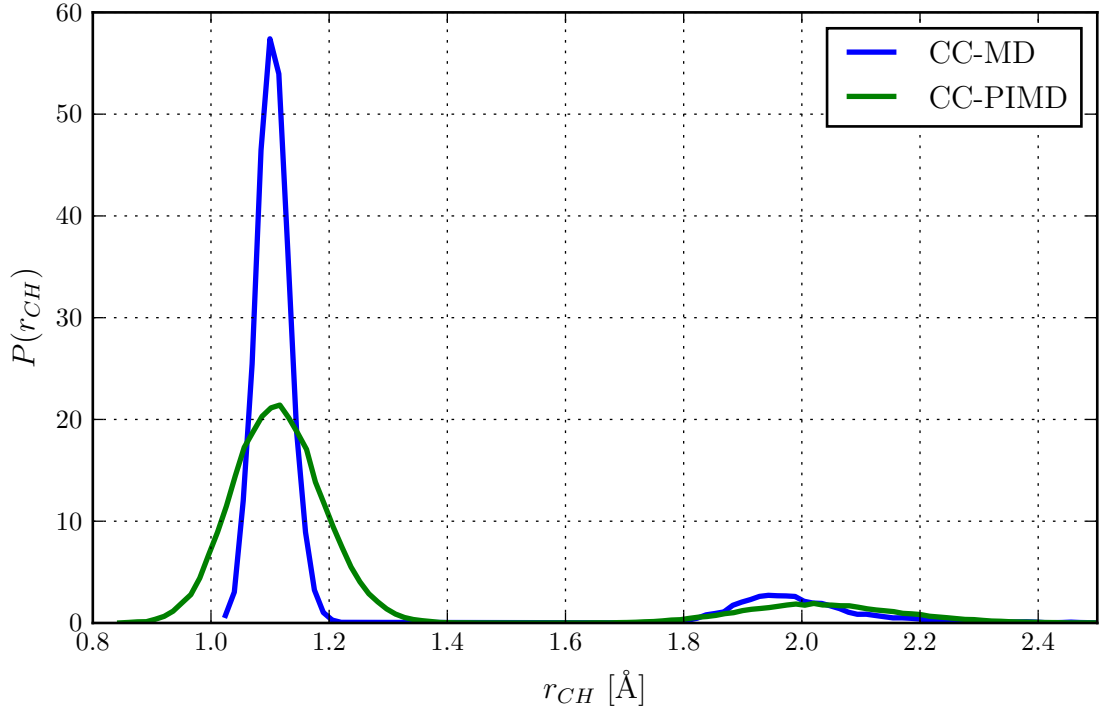}}
        \caption{Partial C--H PCF obtained from the CC-MD and CC-PIMD simulations. The first peak corresponds to the intramolecular C--H bonds, while the broader contribution includes the carbon--shared-proton distance C--H$^+$.}
        \label{fig:ASYM_pcf_ch}
    \end{figure}

The inclusion of NQEs broadens the distributions of bonds involving hydrogen atoms. This is visible in the partial C--H pair correlation function (PCF) in Fig.~\ref{fig:ASYM_pcf_ch}: the first peak is reduced to about one third of its classical height. The mean heavy-atom distances are affected only weakly and remain within the statistical fluctuations, but their distributions respond differently. The O--O distance distribution changes little, whereas the C--O double-bond distribution becomes noticeably broader. The largest heavy-atom angular response is found for the $O_1H^+O_2$ angle, which decreases from 168.2$^\circ$ in CC-MD to 164.9$^\circ$ in CC-PIMD, consistent with the enhanced flexibility of the shared proton.

\begin{figure}[htbp]
        \centering
        \resizebox{0.95\textwidth}{!}{\input{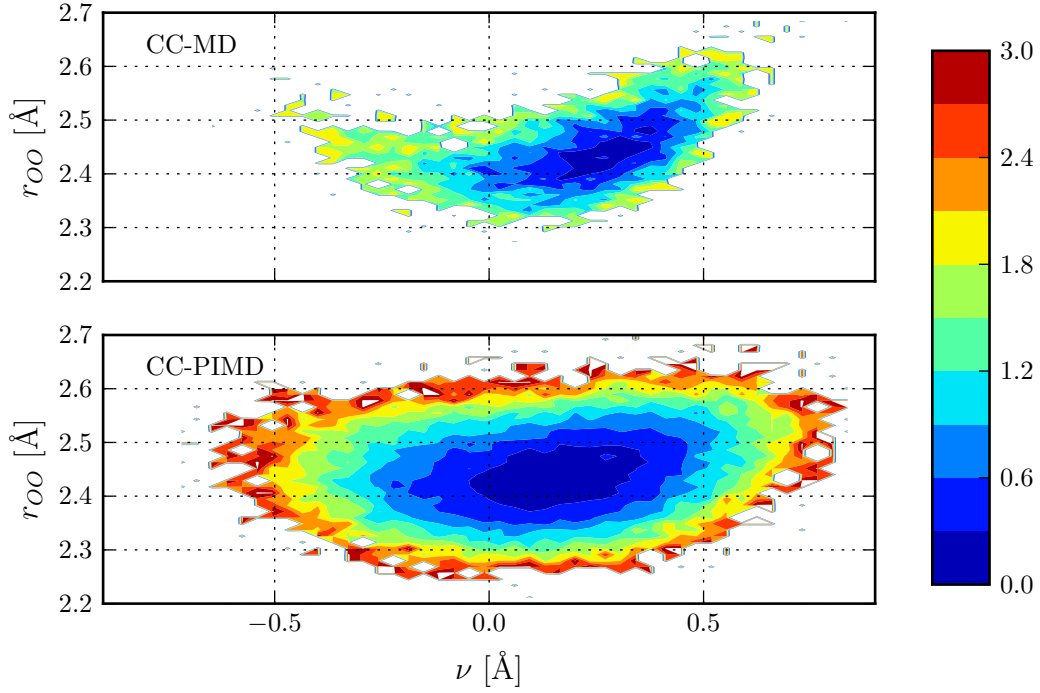}}
        \caption{Free-energy distribution, in kcal~mol$^{-1}$, of the shared proton in our CC-MD and CC-PIMD simulations as a function of the intermolecular O--O distance and the proton reaction coordinate $\nu$.}
        \label{fig:ASYM_zundel_eigen_original_mu}
    \end{figure}

Following previous works~\cite{Tuckerman1997,spura2015ccpimd}, we define the proton reaction coordinate as
\begin{equation}
\nu = r_{O_{\mathrm{H_2O}}H^+} - r_{O_{\mathrm{CH_2O}}H^+},
\end{equation}
so that positive values correspond to configurations in which the proton is closer to formaldehyde. The free-energy distribution along $\nu$ and the intermolecular O--O distance is shown in Fig.~\ref{fig:ASYM_zundel_eigen_original_mu}. In classical CC-MD, the proton is closer to formaldehyde in 81.7\% of the sampled configurations, and the mean reaction coordinate is $\langle\nu\rangle=0.206$~\AA\@. In CC-PIMD, the proton is more delocalized: the formaldehyde-side probability decreases to 61.1\%, and the mean reaction coordinate shifts to $\langle\nu\rangle=0.135$~\AA\@. Thus, NQEs do not remove the asymmetry of the hydrogen bond, but they strongly reduce it.

\begin{figure}[htbp]
        \centering
        \resizebox{0.95\textwidth}{!}{\input{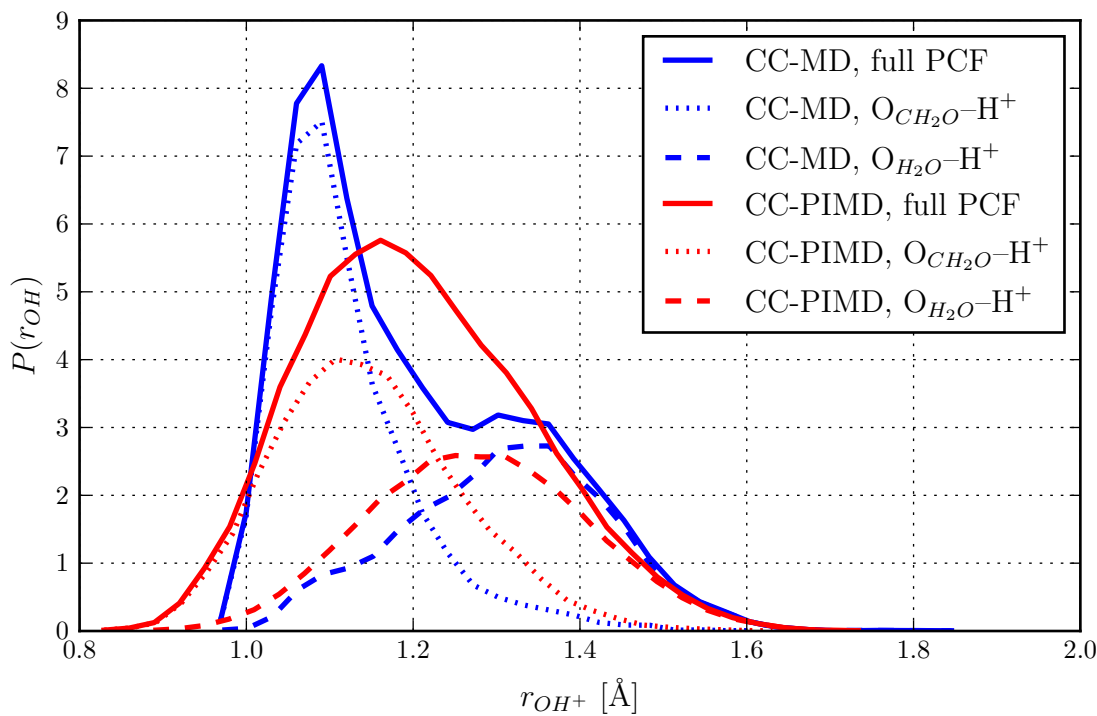}}
        \caption{Partial O--H$^+$ PCF obtained from the CC-MD and CC-PIMD simulations, decomposed into the contributions from the formaldehyde oxygen O$_{\mathrm{CH_2O}}$ and the water oxygen O$_{\mathrm{H_2O}}$ to the shared proton H$^+$.}
        \label{fig:ASYM_pcf_ohplus_detail}
    \end{figure}

The same delocalization is apparent in the decomposition of the partial O--H$^+$ PCF in Fig.~\ref{fig:ASYM_pcf_ohplus_detail}. In classical CC-MD, the total distribution is bimodal, reflecting a proton that is preferentially attached to formaldehyde. In the quantum simulation, the two contributions overlap much more strongly and merge into a broad total peak. The shared proton therefore samples configurations that are largely inaccessible in the classical trajectory.

\subsection{Nuclear magnetic resonance (NMR) response of the asymmetric hydrogen bond}

Nuclear magnetic shielding tensors are highly sensitive to small changes in nuclear geometry and electronic structure. A bond-length change of only 0.0001~\AA\ can shift a computed chemical shielding by about 0.1~ppm~\cite{Grimmer_1993}. Classical vibrational averaging over the Born-Oppenheimer surface can therefore be as important as post-HF electron correlation for quantitative shieldings~\cite{Vaara1998,Dracinsky2009,Ceriotti2013}. PIMD extends such averaging by including zero-point motion, anharmonicity, and tunneling without relying on a perturbative expansion. Previous PIMD studies have shown that nuclear quantum fluctuations can be essential for agreement with experimental NMR observables~\cite{Boehm2001,Schulte2006,Shiga2010,Ceriotti2013,Dracinsky2014}.

The NMR application is a stringent test of the present dynamics because the observable depends on the joint distribution of nuclear configurations and electronic response. For each snapshot, we evaluate the shielding tensor and report the isotropic shielding $\sigma=\mathrm{Tr}(\boldsymbol{\sigma})/3$ as well as the shielding anisotropy $\Delta$. For a fixed reference compound, chemical-shift changes follow the opposite sign convention, $\Delta\delta\simeq-\Delta\sigma$. Thus, a decrease in shielding corresponds to a downfield chemical-shift change. This distinction is useful below because the simulations directly provide shieldings, whereas experimental discussion is usually phrased in chemical shifts.

\begin{figure}[htbp]
        \centering
        \resizebox{0.95\textwidth}{!}{\input{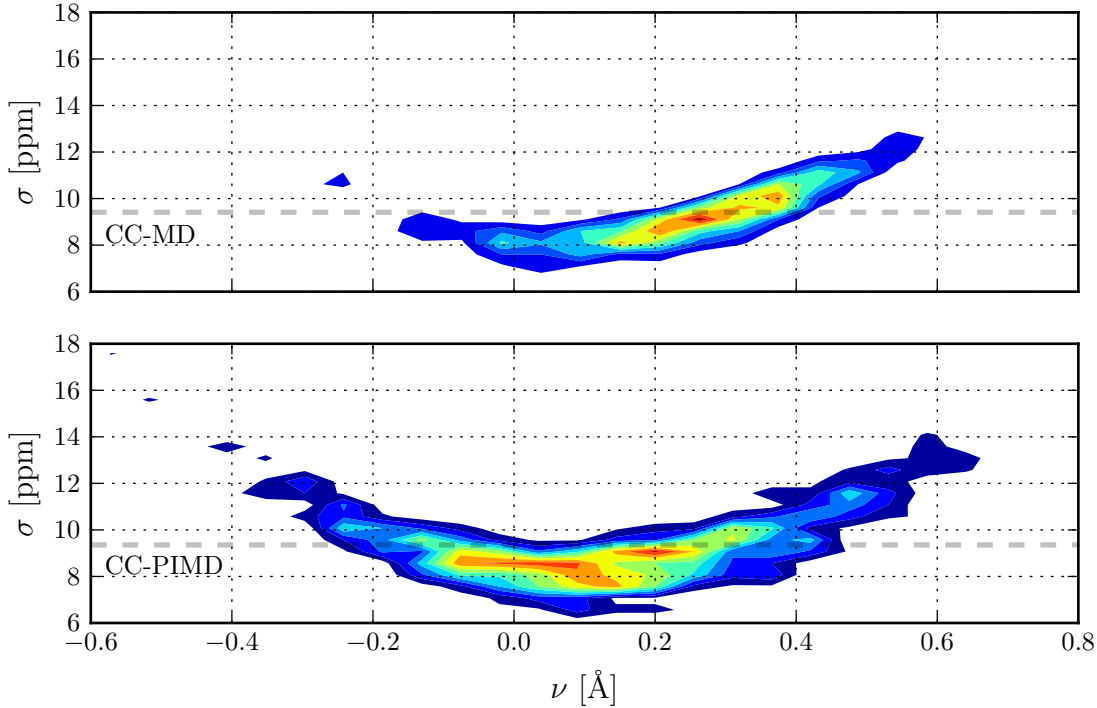}}
        \caption{Distribution of the isotropic nuclear magnetic shielding $\sigma$ of the proton shared by water and formaldehyde, in ppm, as a function of the proton reaction coordinate $\nu$. The horizontal lines denote the average isotropic shieldings.}
        \label{fig:ASYM_isotropic_proton}
    \end{figure}

In Fig.~\ref{fig:ASYM_isotropic_proton}, the isotropic $^1$H shielding of the shared proton is shown as a function of the reaction coordinate $\nu$. Compared with the classical trajectory, the quantum trajectory samples configurations closer to the middle of the hydrogen bond. The PIMD distribution is also more structured in the $\sigma$--$\nu$ plane, as expected for a proton that is dynamically shared between two sites. Although this nonlinear dependence could, by itself, shift the average shielding~\cite{Hassanali2012}, the increased sampling of more shielded configurations is compensated here by the shift of the equilibrium reaction coordinate toward the center of the hydrogen bond. Consequently, the average shared-proton shielding changes by only $-0.05$~ppm at the CC level, even though the underlying proton distribution is strongly modified.

\begin{figure}[htbp]
        \centering
        \resizebox{0.95\textwidth}{!}{\input{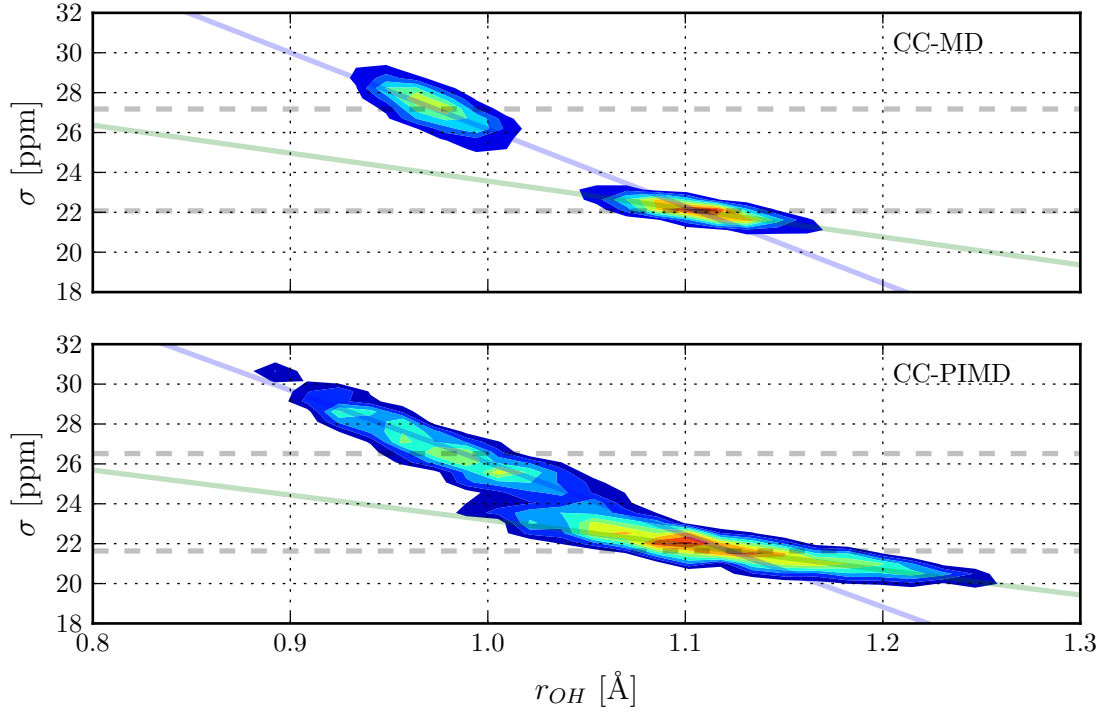}}
        \caption{Distribution of the isotropic shielding of the hydrogens bound to water and carbon. The horizontal lines denote average isotropic shieldings, and the linear fits are included as guides to the eye.}
        \label{fig:shielding_hydrogen}
    \end{figure}

The covalently bound hydrogen atoms show a more direct relation between shielding and bond length. As shown in Fig.~\ref{fig:shielding_hydrogen}, their isotropic shieldings are approximately linear functions of the corresponding X--H distance over the thermally sampled range. NQEs broaden the sampled bond-length distributions, but the resulting average shielding shifts are modest. The NMR sensitivity of the system is therefore not simply controlled by the largest structural fluctuation; it depends on how each nucleus samples the local shielding surface.

\begin{figure}
\centering
\includegraphics[width=0.96\textwidth]{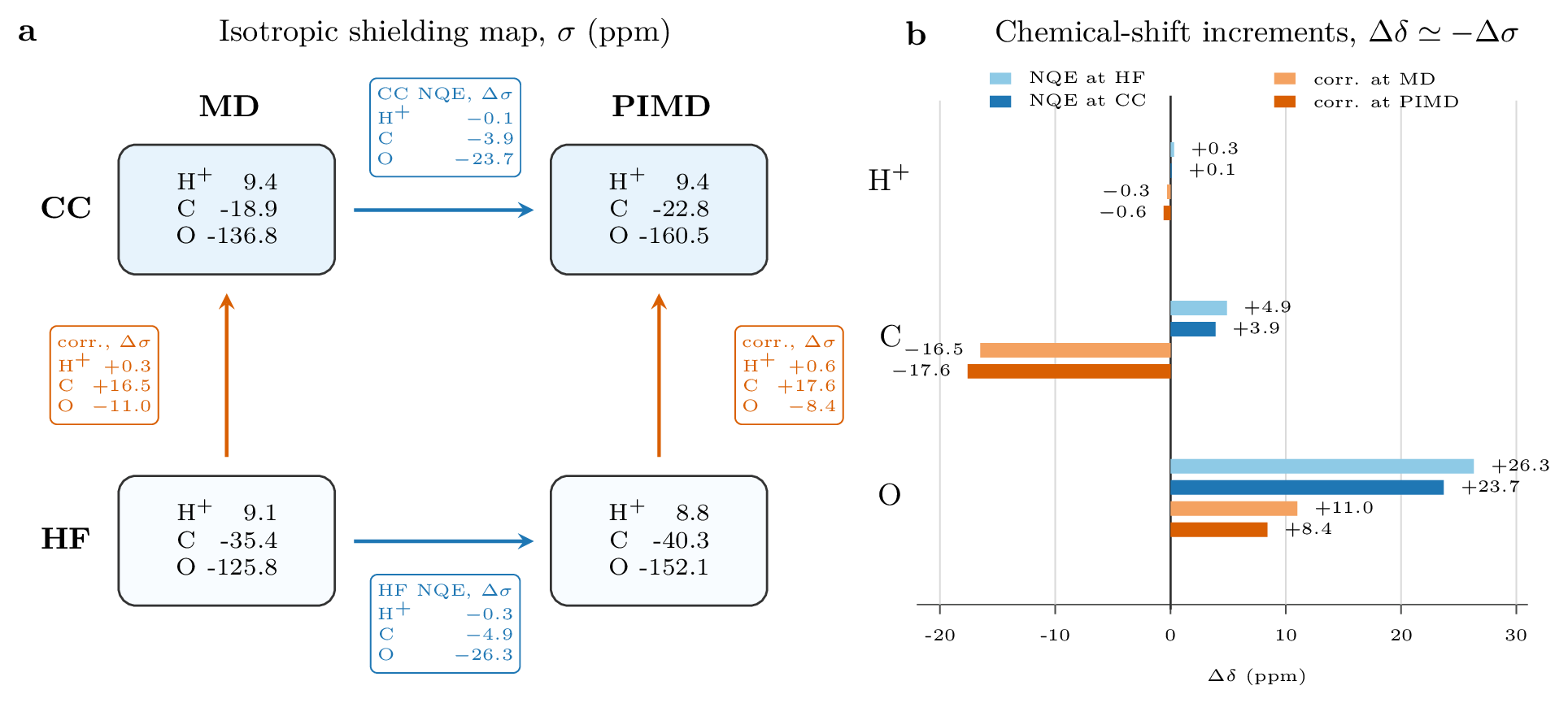}
\caption{Decomposition of the NMR response of the shared proton and the formaldehyde C=O unit. Panel (a) gives ensemble-averaged isotropic shieldings $\sigma$ in ppm at the HF and CC levels for classical nuclei (MD) and quantum nuclei (PIMD). Arrows report the corresponding shielding increments $\Delta\sigma$ due to NQEs or electron correlation. Panel (b) shows the same increments in the chemical-shift convention. For a fixed reference compound, chemical-shift changes follow the opposite sign convention, $\Delta\delta\simeq-\Delta\sigma$, where positive values correspond to downfield shifts.}
\label{fig:nmr_shift_map}
\end{figure}

The shift map in Fig.~\ref{fig:nmr_shift_map} makes the cancellation patterns explicit. At the CC level, NQEs decrease the carbonyl oxygen shielding by 23.7~ppm and the formaldehyde carbon shielding by 3.9~ppm, corresponding to downfield chemical-shift increments of the same magnitudes. The shared proton, in contrast, is almost unchanged in the isotropic average. Electron correlation has a different nuclear fingerprint: at the PIMD geometries, it increases the carbon shielding by 17.6~ppm, decreases the carbonyl oxygen shielding by 8.4~ppm, and increases the shared-proton shielding by 0.6~ppm. Hence, correlation and NQEs act in opposite directions for the carbon chemical shift, but in the same downfield direction for the carbonyl oxygen. The proton shielding is a particularly delicate case because a small isotropic average masks a substantial reshaping of the sampled proton-transfer coordinate.

\begin{figure}[htbp]
        \centering
        \resizebox{0.95\textwidth}{!}{\input{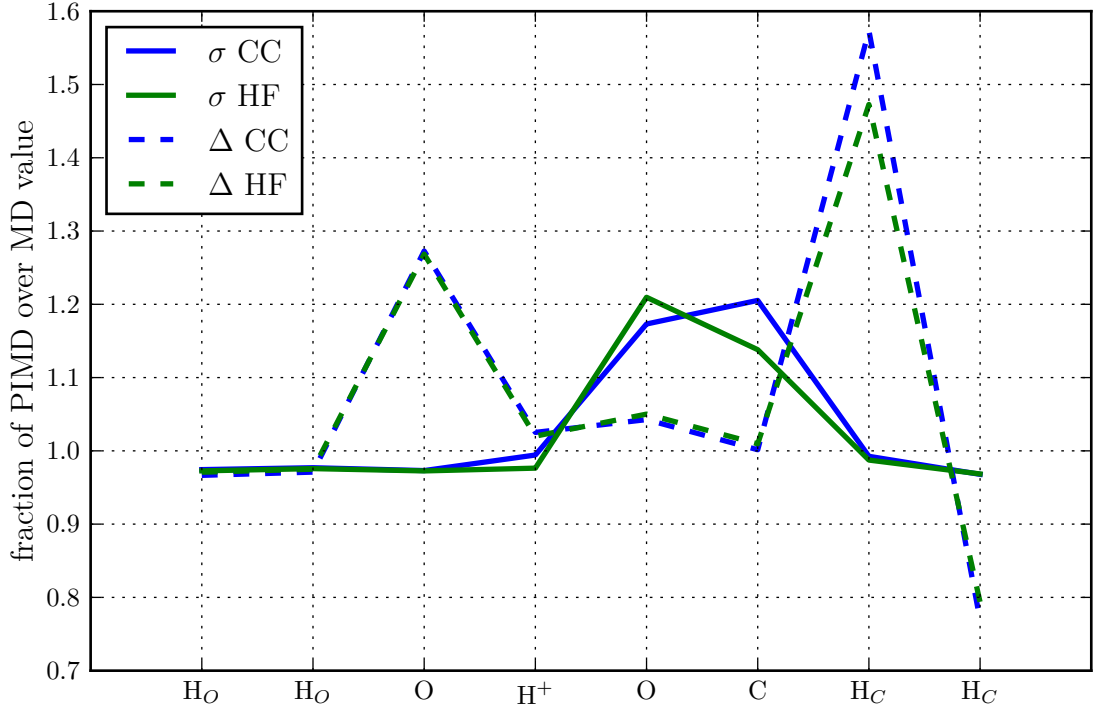}}
        \caption{Impact of nuclear quantum effects on the isotropic nuclear magnetic shielding $\sigma$ and the shielding anisotropy $\Delta$ at different electronic-structure levels. The lines are guides to the eye.}
        \label{fig:ASYM_lot_impact_on_nmr_tensor}
    \end{figure}

Figure~\ref{fig:ASYM_lot_impact_on_nmr_tensor} extends this comparison to all nuclei and includes both $\langle\sigma\rangle$ and $\langle\Delta\rangle$. HF and CC agree qualitatively on the direction of most NQE-induced shifts, but important quantitative differences remain. The largest difference between HF and CC is found for the carbonyl group of formaldehyde, where correlation effects are known to be significant~\cite{Gauss1995a}. For the peripheral hydrogens, both HF-PIMD and CC-PIMD predict a deshielding of about 0.5~ppm. For the shared proton, however, HF-PIMD predicts a similar deshielding while CC-PIMD does not. This cancellation highlights the need to include electron correlation and NQEs simultaneously when quantitative shielding shifts are sought.

The contrast between $\sigma$ and $\Delta$ is also important. For the anisotropy, HF and CC provide a more consistent picture of the NQE response than for the isotropic shielding, with the main difference occurring for the formaldehyde hydrogens. Moreover, NQEs can be much more pronounced in the tensor anisotropy than in the isotropic average. A small change in $\sigma$ can therefore hide compensating but sizeable changes in the principal components of the shielding tensor. This observation is relevant for solid-state NMR and for relaxation measurements, where tensor anisotropies can be directly probed~\cite{Waugh1968,Kukolich1975,HeineNMR2005,ModigHalle2002}.

\begin{figure}[htbp]
        \centering
        \resizebox{0.95\textwidth}{!}{\input{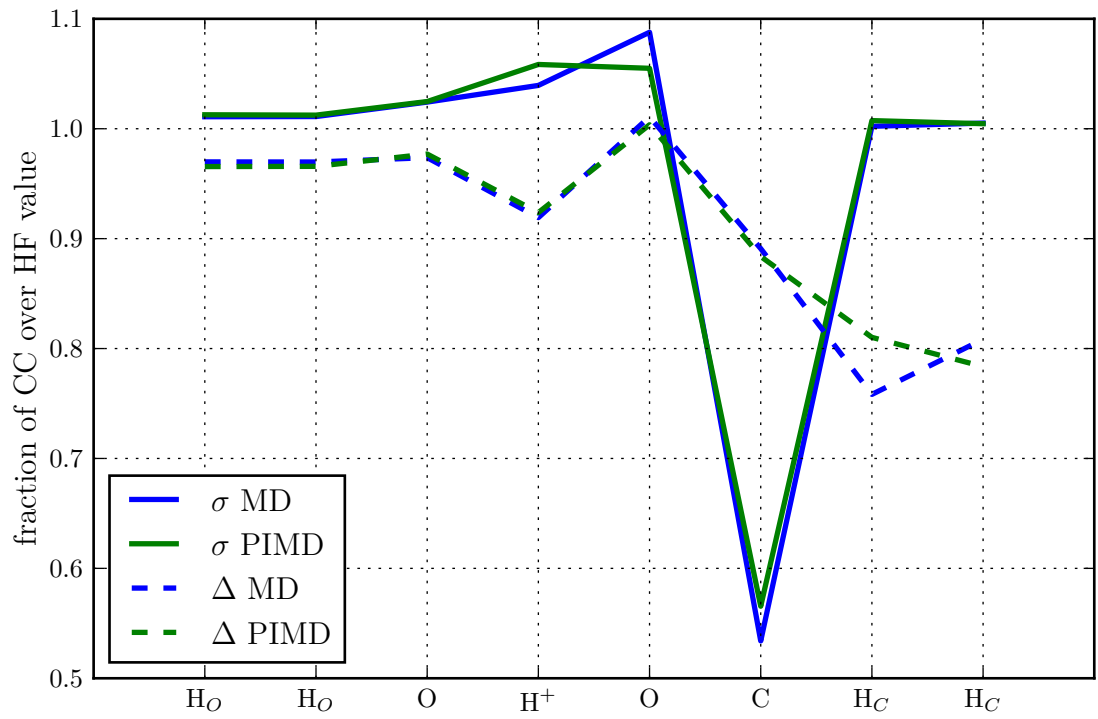}}
        \caption{Impact of electron correlation on the isotropic nuclear magnetic shielding $\sigma$ and the shielding anisotropy $\Delta$ for classical and quantum nuclei. The lines are guides to the eye.}
        \label{fig:ASYM_nqe_impact_on_nmr_tensor}
    \end{figure}

Figure~\ref{fig:ASYM_nqe_impact_on_nmr_tensor} shows the complementary CC/HF ratios of $\langle\sigma\rangle$ and $\langle\Delta\rangle$ for classical and quantum nuclei. The largest electron-correlation effects occur for the shared proton and the formaldehyde unit. The correlation effect on $\sigma$ is not simply correlated with that on $\Delta$, reinforcing that isotropic shieldings alone are not sufficient to characterize the NMR response of an asymmetric hydrogen bond. The absolute CCSD(T) ensemble averages and their snapshot-to-snapshot standard deviations are collected in Table~\ref{tab:nmr_statistics}.

\begin{table}
\caption{Ensemble-averaged NMR shielding parameters from the CC-MD and CC-PIMD trajectories. Values are reported in ppm; the rows labeled SD give the standard deviations over the sampled configurations. The quantities $\lambda_{\max}$ and $\lambda_{\min}$ are the largest and smallest principal shieldings, and $\Omega=\lambda_{\max}-\lambda_{\min}$ is the tensor span. Atom labels $O_{\mathrm{w}}$ and $H_{\mathrm{w}}$ refer to the water fragment, while $O_{\mathrm{f}}$, $C_{\mathrm{f}}$, and $H_{\mathrm{f}}$ refer to formaldehyde.\label{tab:nmr_statistics}}
\centering
\begin{tabular*}{\textwidth}{@{\extracolsep{\fill}}lrrrrrrrr@{}}
\toprule
 & $O_{\mathrm{w}}$ & $H_{\mathrm{w1}}$ & $H_{\mathrm{w2}}$ & $O_{\mathrm{f}}$ & $C_{\mathrm{f}}$ & $H_{\mathrm{f1}}$ & $H_{\mathrm{f2}}$ & $H^+$ \\
\midrule
PIMD $\langle\sigma\rangle$ & 311.975 & 26.491 & 26.545 & -160.537 & -22.799 & 21.481 & 21.785 & 9.357 \\
PIMD SD$(\sigma)$           & 17.016  & 2.585  & 2.621  & 46.757   & 11.169  & 1.112  & 1.099  & 1.700 \\
PIMD $\langle\Delta\rangle$ & 42.893  & 19.783 & 19.862 & 766.068  & 224.929 & 3.921  & 5.378  & 29.492 \\
PIMD SD$(\Delta)$           & 14.522  & 2.293  & 2.309  & 75.579   & 19.458  & 1.428  & 1.744  & 5.137 \\
PIMD $\langle\lambda_{\max}\rangle$ & 340.075 & 39.662 & 39.769 & 349.907 & 127.079 & 23.862 & 25.247 & 28.741 \\
PIMD SD$(\lambda_{\max})$   & 17.600  & 4.012  & 4.069  & 15.172   & 8.623   & 1.722  & 1.931  & 3.641 \\
PIMD $\langle\lambda_{\min}\rangle$ & 280.372 & 19.313 & 19.341 & -543.094 & -112.751 & 19.227 & 18.958 & -3.205 \\
PIMD SD$(\lambda_{\min})$   & 18.741  & 2.041  & 2.079  & 95.664   & 22.465  & 1.547  & 1.323  & 2.809 \\
PIMD $\langle\Omega\rangle$ & 59.703  & 20.349 & 20.428 & 893.001  & 239.830 & 4.634  & 6.289  & 31.946 \\
PIMD SD$(\Omega)$           & 13.832  & 2.180  & 2.178  & 100.421  & 26.203  & 2.301  & 2.466  & 5.364 \\
\midrule
MD $\langle\sigma\rangle$   & 320.605 & 27.187 & 27.170 & -136.845 & -18.915 & 22.192 & 21.949 & 9.410 \\
MD SD$(\sigma)$             & 9.248   & 1.097  & 1.112  & 33.696   & 5.200   & 0.508  & 0.498  & 1.443 \\
MD $\langle\Delta\rangle$   & 33.701  & 20.467 & 20.455 & 734.745  & 224.589 & 5.098  & 3.418  & 28.767 \\
MD SD$(\Delta)$             & 6.643   & 0.939  & 0.946  & 59.594   & 9.985   & 1.056  & 0.600  & 4.759 \\
MD $\langle\lambda_{\max}\rangle$ & 342.641 & 40.817 & 40.791 & 352.864 & 130.770 & 25.549 & 24.056 & 28.344 \\
MD SD$(\lambda_{\max})$     & 11.569  & 1.613  & 1.613  & 9.411    & 4.159   & 1.017  & 0.792  & 2.682 \\
MD $\langle\lambda_{\min}\rangle$ & 290.138 & 19.739 & 19.735 & -497.931 & -106.660 & 19.741 & 20.354 & -3.001 \\
MD SD$(\lambda_{\min})$     & 9.823   & 0.895  & 0.918  & 76.354   & 12.287  & 0.714  & 0.731  & 2.853 \\
MD $\langle\Omega\rangle$   & 52.503  & 21.078 & 21.056 & 850.796  & 237.430 & 5.808  & 3.702  & 31.345 \\
MD SD$(\Omega)$             & 7.627   & 0.975  & 0.974  & 82.669   & 14.306  & 1.340  & 0.971  & 4.947 \\
\bottomrule
\end{tabular*}
\end{table}

\clearpage

\section{Conclusions}

We have presented an accelerated CC-PIMD method for simulations in which electron correlation and nuclear quantum effects are treated in an on-the-fly manner. The method is organized around qRPC, which provides a natural CCSD/HF decomposition: HF is the inexpensive auxiliary potential evaluated on the full ring-polymer, while the expensive CCSD--HF residual is evaluated on a contracted ring-polymer. This imaginary-time acceleration is combined with two second-generation Car-Parrinello-like real-time predictors: the HF reference is propagated using a density-matrix predictor, and the CCSD and $\hat{\Lambda}$ amplitudes are transformed to a smooth orthogonal-basis representation, extrapolated in time, and transformed back to the instantaneous molecular orbital basis. The combination of all three acceleration layers is essential for making such correlated PIMD trajectories practical.

For the water--formaldehyde shared proton, the amplitude predictor reduces the number of CCSD and $\hat{\Lambda}$ iterations by about 30\%, and a two-iteration cluster approximation reproduces fully converged energies and forces with very small errors along the tested trajectory. Nuclear quantum effects strongly reduce the asymmetry of the shared proton, shifting the mean proton-transfer coordinate from 0.206 to 0.135~\AA\ and lowering the probability of finding the proton closer to formaldehyde from 81.7\% to 61.1\%.

The NMR results show that nuclear quantum effects and electron correlation are comparable in magnitude and may partially cancel. This is especially clear for the shared proton and for the carbonyl group of formaldehyde. The shielding anisotropy can respond more strongly than the isotropic shielding, which cautions against drawing conclusions from isotropic averages alone. Overall, the results demonstrate that asymmetric hydrogen bonds require a simultaneous and dynamically consistent treatment of correlated electronic structure and nuclear quantum fluctuations.

\section*{Acknowledgements}

The authors gratefully acknowledge the computing time made available to them on the high-performance computer Noctua at the NHR Center Paderborn Center for Parallel Computing (PC2).
This center is jointly supported by the Federal Ministry of Research, Technology and Space and the state governments participating in the National High-Performance Computing (NHR) joint funding program (www.nhr-verein.de/en/our-partners).
Part of the research was funded by the DFG (project numbers 417590517/CRC1415 and 519869949).

\bibliography{citations}

\end{document}